\documentclass[preprint]{aastex} 
\usepackage{graphicx}
\usepackage{txfonts}
\usepackage{natbib} 
\usepackage{fullpage}
\bibpunct{(}{)}{;}{a}{}{,} 
\begin{document}
%
   \title{Sun as a star observation of White-Light Flares}


   \author{M. Kretzschmar}
   \affil{LPC2E, UMR 6115 CNRS and University of Orl\'eans,
              3a av. de la recherche scientifique, 45071 Orléans, France\\}
    \email{matthieu.kretzschmar@cnrs-orleans.fr}



 
\begin{abstract}
  Solar flares radiates energy at all wavelengths, but the spectral distribution of this energy is still poorly known. White-light continuum emission is sometimes observed and the flares are then termed "white-light flares" (WLF).
   In this paper, we investigate if all flares are white-light flares and how is the radiated energy spectrally distributed.
   We perform a superposed epoch analysis of spectral and total irradiance measurements obtained since 1996 by the SOHO and GOES spacecrafts at various wavelength, from Soft X-ray to the visible domain.
   The long-term record of solar irradiance and excellent duty cycle of the measurements allow us to detect a signal in visible irradiance even for moderate (C-class) flares, mainly during the impulsive phase. We identify this signal as continuum emission emitted by white-light flares, and find that it is consistent with a blackbody emission at $\sim$9000K. We estimate for several sets of flares the contribution of the WL continuum and find it to be of $\sim 70$ \% of the total radiated energy. We re-analyse the X17 flare that occurred on 28 October 2003 and find similar results.
   This paper brings evidence that all flares are white-light flares and that the white-light continuum is the main contributor to the total radiated energy; this continuum is consistent with blackbody spectrum at $\sim$9000K. These observational results are important in order to understand the physical mechanisms during flares and open the way to a possible contribution of flares to TSI variations.
 \end{abstract}
%

\section{Introduction}
Solar flares are huge bursts of energy in the atmosphere of the Sun. 
Satellite observations have emphasized their observations in the extreme ultraviolet (EUV) and Soft X-rays (SXR) domain, which were not possible before the space age. At these wavelengths, the emission increases drastically and can stand for a few hours (depending on the flare magnitude). Flares however were first observed from the ground in the visible domain \citep{1859MNRAS..20...13C,1966SSRv....5..388S,Neidig:1989ft}
and the advent of spectroscopic observations revealed the increase of solar flux in visible spectral lines, sometimes changing from absorption to a emission profile \citep{Hale:1931lr,1998SoPh..182..447S}. \\
 Flare emission in the visible domain is thus well known to occur in visible chromospheric lines like the Balmer and Ca II K lines \citep{1990ApJ...363..318C,1992A&A...256..255F,1994SoPh..152..393H}. There is, however, another -and less understood- contribution to the visible emission that is due to the enhancement of the continuum. The term 'white-light' (WL) is used to refer to visible continuum enhancement. The study of WL flares have been rendered very difficult by their short duration and very low contrast which makes their observations from Earth rare and of poor quality. \cite{Neidig:1989ft} emphasized this aspect together with the suggestion that WL emission can be quite common and can constitute up to 90$\%$ of the total energy radiated by the flare. This has, however, never been confirmed and several questions remain about WL emission, like: What is the contribution of WL emission to the total flare energy ? Does it appear only in very large flares ?  How and where is it produced ? \\
There are only a few observations of WLFs in space as the available measurements do not in general have the required high spatial and temporal resolution -as well as duty cycle-. Several WLFs have however been observed with Yohkoh \citep[e.g.][]{Matthews:2003fk} and TRACE  \citep[e.g.][]{Hudson:2006aa}; these studies have confirmed the importance of WLFs and their association with the hard X-rays in the impulsive phase of the flare. Furthermore, many flares were observed with no associated white-light emission and with no means to determine if this was due to instrumental limitations or to the actual absence of WL emission. Recent studies \citep{2008ApJ...688L.119J,1674-4527-9-2-001} identified WL emission in relatively small flares, which support the view that WL emission could be a peculiar feature of flare and not only associated to the largest flares. \\

Knowing how much energy is radiated as a whole and at each wavelength is important both to understand the physical processes in the solar atmosphere and the impact of flares on Earth. Ideally, it would however be necessary to observe the Sun at all wavelengths with sufficient contrast (implying good spatial and temporal resolution) and duty cycle (to avoid missing the flare). One must often deal with partial observations only, especially at ultraviolet and visible light where contrast is weak and space instrumentation is rare. This explains why we still have a partial view of how the flare energy is spectrally distributed. \cite{2004GeoRL..3110802W,Woods:2006aa} provided the first direct observation of the flare signal in the total solar irradiance (TSI), allowing thus a precise estimate of the the total energy radiated by this extremely large (X17) flare of October 28, 2003. Using additional solar irradiance observations in the XUV (shorter than 27nm) and the FISM flare model up to 190nm, they concluded that about half of the energy is radiated below 190nm, most of it coming from wavelengths shorter than 14nm and the other half coming thus from near UV, visible and infrared light. These important results for very large flares suffer, however, from the absence of direct observations at wavelengths longer than 27nm and assume that the TSI time profile follows the flare time profile in XUV. \\
In this paper, we show that white-light continuum is commonly produced in basically all flares and that it represents most of the flare energy. In section \ref{sec_data}, we present the data and the analysis method. We follow \cite{2010NatPh...6..690K} and perform a superposed epoch analysis of solar \emph{irradiance} (i.e. integrated solar flux) in the visible domain during flares. The exceptional duty cycle and long-term monitoring of the Sun Photometer (SPM) instrument counterbalance the absence of spatial resolution and the presence of background temporal fluctuations due to p modes, which allows to reveal the flare signal at visible wavelengths and to identify it as WL continuum (section \ref{sec_res}). In section \ref{sec_discuss}, we estimate the spectral distribution of the radiated energy for average flares of different amplitude as well as for the single X17 flare of October 28, 2003, for which we present new observations. We conclude in section \ref{sec_conclu}.
%
%
%
\begin{figure*}[!ht] 
    \begin{center}
  \includegraphics[width=0.8\textwidth,height=0.45\textheight]{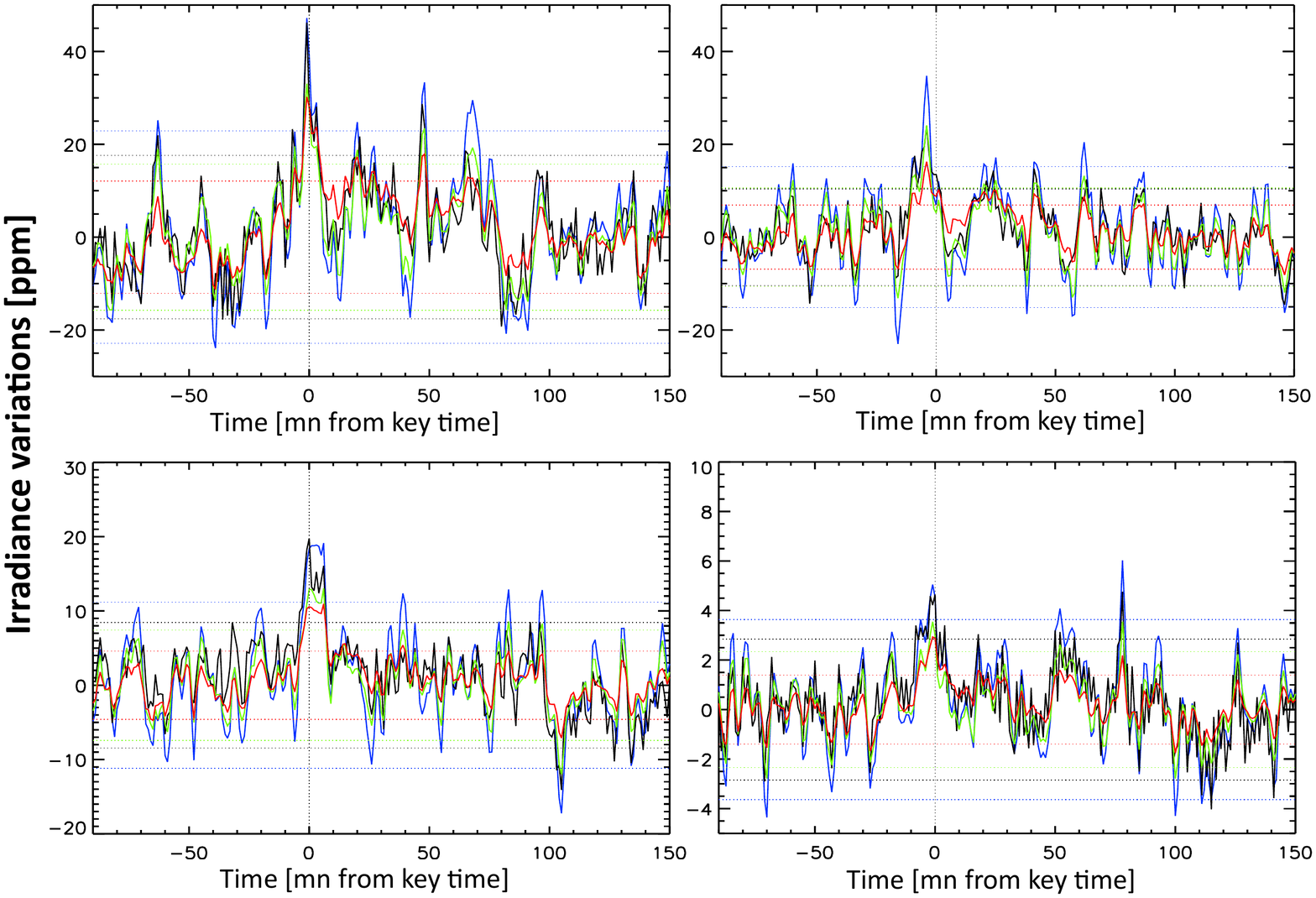}
    \end{center}
    \caption{{\bf Average flare light curves for TSI (black), and the three visible SPM channels (blue, green, and red)}. \textit{Upper left panel}: average over 43 flares from X17.2 to X1.3. \textit{Upper right panel}: average over 68 flares from X1.3 to M6.4. \textit{Lower left panel}: average over 140 flares from M6.4 to M2.8. \textit{Lower right panel}: average over 1850 flares from M2.8 to C4.\label{fig_res}. Horizontal lines show the $\pm 2\sigma$ limits computed from -90 to 150 minutes excluding 60min centered around the peak time.}
\end{figure*}
%
%
%
%
\section{Data and Analysis \label{sec_data}}
%
%
The data we used in this study are full Sun fluxes (i.e. solar irradiance) observed by the SOHO and GOES spacecrafts: The \textit{Total Solar Irradiance (TSI)} (solar flux integrated over all wavelengths), measured by the VIRGO/PMOv6 instrument onboard SOHO; three \textit{Visible Solar Irradiance} from the VIRGO/SPM instrument (still SOHO) consisting of passbands of 5nm respectively centered on 402nm (blue), 500nm (green), and 862nm (red).
These irradiance time series cover the period from 1996 to mid 2008 with a 1-minute time step with very few data gaps. They have been corrected for degradations and can be downloaded from the VIRGO ftp website\footnote{ftp.pmodwrc.ch$/$pub$/$data$/$irradiance$/$virgo$/$1$-$minute\_Data$/$}. Additionally we use the Extreme Ultraviolet (EUV) irradiance in the ranges 0.1-50nm and 26-34nm measured by SOHO/SEM \citep{1998SoPh..177..161J} and the Soft X-ray (SXR) irradiance measured by the GOES satellites.\\
%
The superposed epoch analysis is relatively simple and consists in superposing several time series in which a similar event (here a flare) occurs at the same time. It is useful when the signal of the event is faint with respect to the random (incoherent) background fluctuations. The resulting superposed (or average) curve exhibits thus smaller background fluctuation (typically attenuated by a factor $1/\sqrt{n}$, $n$ being the number of samples) but a stronger signal at the time of the event, which results from the superposition of all the coherent faint signals in each individual time series. This technique has been used to detect flares in total solar irradiance by \cite{2010NatPh...6..690K} and we extend here this work by looking at visible, EUV, and SXR  irradiance time series. \\ 
 The events we considered here are flares observed in Soft X-ray by the GOES satellites from 1996 to 2007, and occurring at heliocentric angle $\theta \le 60\deg$ because at higher $\theta$ the visible emission is likely to be strongly absorbed.
Before superposing the light curves for the TSI and visible irradiance, we subtracted a linear fit from the time series in order to eliminate trends. This in principle could hide meaningful trends (for example linked to the evolution of the active region pre- or post- flare) but it allows us to reach a better signal-to-noise ratio and the analysis without trend removal did not reveal clear results. In section \ref{sec_discuss} we analyse the spectral distribution of the flare energy; for this purpose, we convert the TSI and visible light curves from ppm to irradiance units (W.m$^{-2}$) by simply multiplying the average time series by their average value (1365 W.m$^{-2}$, for the TSI, 1.67W.m$^{-2}$, 1.83W.m$^{-2}$, 0.97W.m$^{-2}$ respectively for the blue, green, and red channel); small departures from theses values (for example by taking 1361W.m$^{-2}$ instead of 1365W.m$^{-2}$ for the TSI) do not affect the result because of the difference of many orders of magnitude between the quiet irradiance and the flare flux. The averaged light curves in physical units for the EUV and soft X-rays channels are obtained by subtracting the background of each time series before averaging.
%
%
\section{White-light Flare observations in Sun-as-a-star measurements \label{sec_res}}
Fig.\ref{fig_res} shows the average flare light curves for 4 sets of flares of decreasing magnitude in the visible and TSI channels. The GOES SXR peak time has been chosen as the key time for each flare. The first immediate result is the appearance in all channels of a peak around the key time that is caused by the flares. Importantly, this is true down to C-class flare, although it can be noted that the signal-to-noise ratio is smaller. This shows in particular that visible emission is ubiquitous in flares.  We can note that other peaks exceed the 2$\sigma$ limits (about 5 points over 100, as expected for random noise); they are, however, less pronounced and appear at different times for the different flare sets. The curves also display oscillations with roughly a 10 to 20-minute period, whose amplitude, however, is below the 2$\sigma$ level (see also fig.\ref{fig_spec0_2100}, before peak time); it is not clear whether these oscillations can be related to actual solar processes (pre- and post- flare accoustic waves) or if they are artifacts of the instrumental noise. The fact that they are not in phase for different flare sets rather points towards an instrumental effect.\\
The curves exhibit several additional interesting features. \\
\begin{figure}[!h] 
    \begin{center}
  \includegraphics[width=0.4\textwidth,height=0.22\textheight]{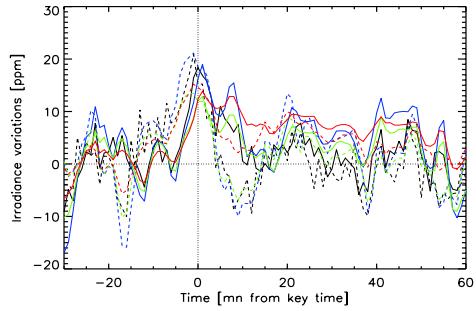}
    \end{center}
    \caption{{\bf Influence of the key time on the analysis}. Dashed lines are average flare time series using the GOES 0.1-0.8 nm flux peak time, thick lines are average flare time series using the peak time of the derivative of the GOES 0.1-0.8 nm flux. Black is for TSI, red for 862nm, blue for 402nm, and green for 500nm. The average is made over the 150 largest flares in our sample, from X28 to M5. \label{fig_peaktime}}
\end{figure}
\begin{figure}[!h] 
    \begin{center}
  \includegraphics[width=0.4\textwidth,height=0.22\textheight]{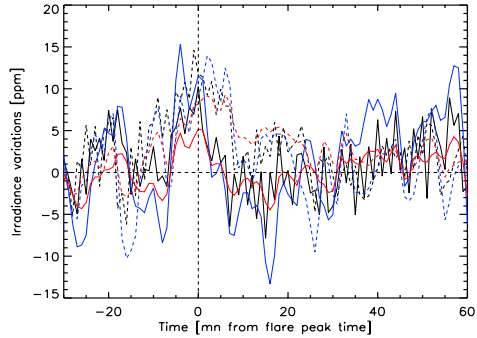}
    \end{center}
    \caption{{\bf Influence of the impulsive phase duration on the analysis}. Thick lines show average time series for flares with short impulsive phase (equivalent X-ray class M2), dashed lines for long impulsive phase (equivalent X-ray class M5). SPM green is not shown for clarity but behaves similarly to SPM blue. \label{fig_duration}}
\end{figure}
First, most of the visible emission belongs to the impulsive phase of the flare. This can be better seen on  fig.\ref{fig_peaktime}: 
when the key time is the GOES SXR peak time, the maximum of the curves precedes it, while when the key time is the peak of the derivative of the SXR flux, the maximum occurs at t=0. The amplitude of the increase, however, is larger when using the SXR flux peak; we attribute this to the fact that the Neupert effect does not hold rigorously in all flares.  Fig. \ref{fig_duration} shows in a similar way the effect of the duration of the impulsive phase, computed as the duration of the rising phase of the SXR flare, i.e. $t_{SXR}^{peak} -t_{SXR}^{start}$. The flare signal in the visible channels and in the TSI nearly disappears after the SXR peak time for flares with short impulsive phase ($\le 6$min), while it persists for flares with longer impulsive phase ($> 6$min). This suggests that visible emission is present in both compact and gradual flare, but that it also persists for some time during the gradual phase of gradual flares.\\
%
%
The second point to note on fig.\ref{fig_res} is the longer decay of the red channel after the peak, which can be attributed to the chromospheric Ca II line emission that is included in this channel. This constitutes a clear signature of the gradual phase, that is not seen in the blue and green channel, and is much less obvious in the TSI measurements. \\
Thirdly, the flare emission in the blue tends to be larger than in the two other visible channels. We come back to this point below, but let's note now that it agrees with the general statement that stellar flares are "blue", i.e. display more emission towards the UV \citep[e.g.][]{2003ApJ...597..535H}. \\
Last, but not least, the relative increase in the visible channels is nearly of the same amplitude as for the TSI. This strongly suggests that the visible emission has a dominant contribution in the total energy radiated by flares. We also come back to this point later. \\

%
%
Fig.\ref{fig_res} shows that the SPM visible emission is observed on average in all flares down to C-class ones. Does this emission come from lines or continuum ? The SPM channels have been chosen to be in the continuum of the solar spectrum and their response functions computed for the quiet Sun \citep{2005ApJ...623.1215J} shows that the the blue and green channel corresponds to the deep photosphere (near $\tau _{500}=1$), while the red channel has a larger contribution of the upper photosphere (the Ca II line).
Since, however, a multitude of lines are present all over the optical solar spectrum, we have looked at the modeled spectrum from \cite{Chance:2010lr}, the observed spectrum by \cite{2004AdSpR..34..256T} and the observed identified line lists by \cite{Allende-Prieto:1998qy}. The blue and green channels include only a few weak lines from neutrals (mainly Fe I), which strongly suggests that the increase of emission, observed during the impulsive phase, is caused by a general increase of the continuum level.  The next section shows that this continuum is furthermore consistent with a blackbody spectrum at T$\sim$9000K, which agrees with solar and stellar observations of WLFs. \\
 Since we access only light curves averaged over several flares, we can wonder how often this WL continuum emission does actually occur. To test this, we have substituted in our sample some of the time series that include flares, by time series that do not include them (randomly selected at time where no flares occur). The flare signal for the less energetic flare set (bottom right panel in fig.\ref{fig_res}) becomes visible when only one out of five time series is replaced. This indicates roughly that at least 80\% of the flares have WL continuum emission and can thus be considered as WLFs. \\

\begin{figure*}[!ht] 
    \begin{center}
  \includegraphics[width=0.7\textwidth]{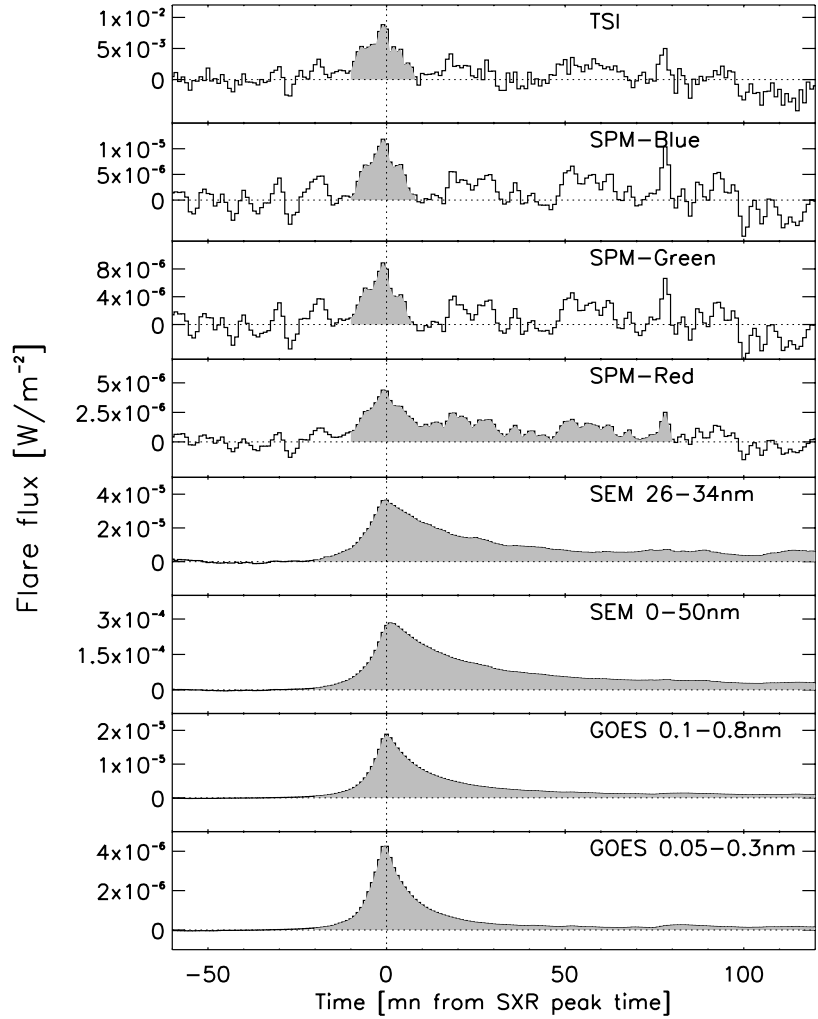}
    \end{center}
    \caption{{\bf Flare light curves averaged over 2100 flares} (from X17.2 to C4) in various spectral ranges.\label{fig_spec0_2100}}
\end{figure*}
\section{Spectral distribution of flare energy \label{sec_discuss}}
%
\begin{table*}[!t]
\caption{\label{tab_nrj1} {\bf Spectral distribution of flare energy.} From first row to the last one the average corresponds to the five sets of flares shown in Figs.\ref{fig_res} and \ref{fig_spec0_2100}. The two ratio in column 3 to 5 correspond respectively to integrations over all the flare duration and limited to the TSI flare signal period. The three last columns show mean blackbody and flare parameters that can explain the visible emission at the flare peak time.}
\centering
\begin{tabular}{llcccccc}
\hline\hline
 Mean &  Total Energy &Ratio&Ratio&Ratio & T$_{bb}$ & S$_{f}$ & Ratio\\
 X-ray class & TSI (Ergs)&26-34nm/TSI&0-50nm/TSI&0.1-0.8nm/TSI & ($^{\circ}$K) & (arcsec$^{2}$) & Continuum/TSI\\
\hline
X3.2 & 5.9 10$^{31}$           & 0.9\% - 0.8\% & 12\% - 9\% & 1.2\% - 1\% & 9345 & 16.7 & 67\% \\
M9.1 & 1.6 10$^{31}$           & 1.7\% - 0.4 \%  & 23\% - 5\% & 1.0\% - 0.4\%   & 8993 & 13.2 & 85\% \\
M4.2 & 1.3 10$^{31}$ 	& 2.2\% - 0.5\% & 18\% - 6\% & 0.6\% - 0.3\% & 9244 & 7.3 & 74\% \\
C8.7 & 3.6 10$^{30}$ 	 & 1.5\% - 0.5\% & 16\% - 5\% & 0.4\% - 0.2\% & 8655 & 2.4 & 72\%\\
\hline
M2.0 & 5.1 10$^{30}$  & 1.7\% - 0.6\% & 18\% - 6\% & 0.7\% -  0.4\% &  8941K & 2.8 & 69\%\\
\hline
\hline
\end{tabular}
\end{table*}
%
%
%
Fig.\ref{fig_spec0_2100} shows flare light curves in several spectral bands from Soft X-ray to visible and averaged over 2100 flares ranging from X-class to C-class. The EUV and SXR bands show a long gradual phase that can also be seen in the visible red channel that contains a chromospheric contribution. The TSI light curve should also exhibit the gradual phase that occurs at all chromospheric and coronal wavelengths but it does not. The most probable explanation is that the gradual phase is below the noise level, which emphasizes the predominance of the impulsive phase. Because of this fact, we cannot exclude that the blue and green channels also have a gradual phase although smaller than the red channel. We use fig.\ref{fig_spec0_2100} to estimate the spectral distribution of the flare energy and we repeat this for the four sets of flares shown on fig.\ref{fig_res}. Because the gradual phase is hardly observed in the TSI average light curves, we computed the energy radiated in each passband both over the total duration of the flares, i.e. including the long gradual phase when it is present, and over the restricted time for which the flare signal is seen in the TSI. For the visible irradiance and the TSI (EUV and SXR channels respectively) we limit the start of the flare integration at 10min (20min resp.) before the key time, in order to avoid signals potentially due to background fluctuations in the visible and TSI light curves.\\
 Emission excess observed at Earth can be converted in energy release on the Sun with the factor $d_{AU}^2 \times f$ where $f$ takes into account the angular distribution of the emission. We use the following factors: $f=2\pi$ for optically thin emission (0-50nm and 0.1-0.8nm), and $f=1.4\pi$ for the TSI and 26-34nm channel, as suggested by \cite{Woods:2006aa}; assuming the same angular distribution for all channels decreases the optically thin contribution by a factor 1.43. The resulted flare energy distribution is shown in tab.\ref{tab_nrj1} for the two integrations (next section explains the three last columns); it reveals how the short wavelength passbands are a minor contributor to the total energy, even when the gradual phase is not taken into account  only for the TSI. The GOES SXR energy is less than 1\% of the total energy, while all wavelengths below 50nm represent between 10\% and 20\%. The visible and near UV part of the flare spectrum must thus constitute the bulk of the flare energy.
%
%
%
%

\subsection{WL energy}
%
It is difficult to reproduce the exact shape of the continuum spectrum from the observations of only three passbands. However we can make simple considerations and try to estimate the total energy that goes in the continuum.
%
%
%
   \begin{figure}[t] 
   \centering
   \includegraphics[width=0.35\textwidth,height=0.19\textheight]{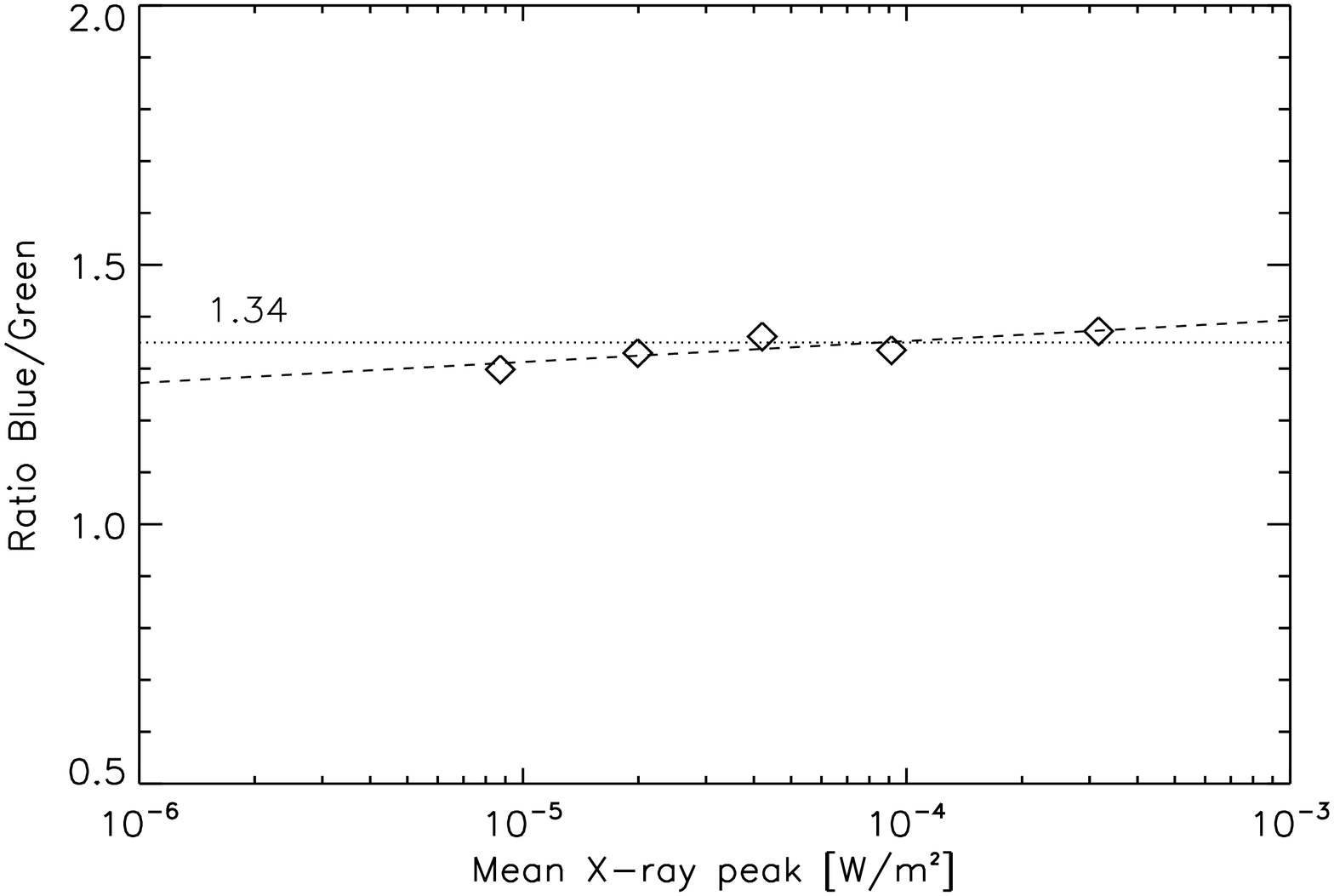}

      \caption{{\bf Ratio between the flare maximum emission in the blue and green channel versus the mean SXR flux peak.} Dotted line is the average ratio of 1.34 and the dashed line is the best fit obtained with the following expression $I^{f}_{Blue}/I^{f}_{Green}=0.04\times\log(E_{SXR})+1.51$. The mean ratio can be reproduced by assuming the flare emission follow a blackbody curve at $\sim9100$K.  }
         \label{Fig_ratioBlueGreen}
   \end{figure}
WLFs are usually divided in two types according to whether the emission exhibits or not a jump at the Balmer and Paschen edges; in our case, the blue and green passbands are just above the Balmer edge and the red channel is just above the Paschen edge; the Hydrogen free-bound emission can thus difficultly explain the observed emission (which does not mean that this emission is not present). Fig.\ref{Fig_ratioBlueGreen} shows the ratio of the flare emission in the blue and green channel $I_{blue}/I_{green}$ at peak time; this ratio is about 1.34 and changes slowly for different flare amplitudes. We can easily convert the ratio values in blackbody temperature; a linear extrapolation of the observed increasing ratio with flare amplitude leads to temperatures ranging from T$\sim$ 8430K for C1 flares to T$\sim$ 9560K for X10 flares.  The slow change in temperature with the SXR flare magnitude indicates that the spectrum shape is similar for different flare magnitudes, the amplitude of the WL continuum depending mainly on the flaring area. Table \ref{tab_nrj1} gives the exact temperature found for each set of flares.
These blackbody temperatures are in good agreement with what has been found previously. \cite{2003ApJ...597..535H} explained the continuum emission of a flare observed on the M dwarf AD Leo with a blackbody temperature near 10000K; \cite{2007ApJ...656.1187F} found a very upper limit of $2.5\times 10^{4}$K to reproduce the spectral ratio between UV and visible emission in several solar WL flares, while recently \cite{Kowalski:2010fj} have provided evidence that the flare emission observed on a dMe4.5e star is composed of a Balmer component at wavelengths shorter than $\sim 380$nm and of a $T \sim 10^{4}$K blackbody component above 400nm.  WL continuum emission is thought to occur in the minimum temperature region or below, where heating is provided by radiative backwarming from the chromosphere as has been found in simulations \citep{2005ApJ...630..573A,2006ApJ...644..484A,Cheng:2010qy}. As a consequence of the very disturbed and unknown state of the flaring atmosphere, the exact processes at play remain unidentified and it is possible that an unknown mechanism produces a flare spectrum that looks similar to a blackbody spectrum near 9000K. 
As supported by our results and the other findings cited above, and in order to compute the flare continuum energy, we assume in the following that the WL continuum follows a blackbody spectrum, keeping thus in mind that its origin is unclear, but that it is supported by various observations. Hydrogen free-bound continua should also be present but are unlike to contribute significantly in the SPM passbands.\\

Assuming that the flare signal in the blue and green channels come from blackbody radiation, we can make simple estimates; we first concentrate on the "average" M2 flare shown on fig.\ref{fig_spec0_2100}. The ratio $I_{blue}/I_{green}$ leads to a blackbody temperature of T$\sim$ 8941K (see table \ref{tab_nrj1}). Multiplying the corresponding blackbody spectrum by the SPM blue channel response and matching the observed value at peak time,
 we obtain a flaring area on the solar surface of 2.8 $arcsec^{2}$. Integrating the blackbody spectrum with this flaring area leads to a total continuum emission at peak time of 6.2 10$^{-3}$W.m$^{-2}$, i.e. about  $\sim$70\% of the total radiated energy at peak time (9 10$^{-3}$W.m$^{-2}$ for the TSI).  We stress here that the estimation of the total energy contained in the WL continuum is independent of the one deduced from the TSI observations.
Doing the same exercise for the total energy radiated over time where the TSI flare profile is available, we find a mean flaring area of 1.4 $arcsec^{2}$ corresponding to a total continuum energy of 3.4 10$^{30}$erg to be compared with the observed total 5.2 10$^{30}$erg, i.e. 60\%. 6\%  comes from wavelengths below 50nm; the remaining could come from EUV wavelengths above 50nm and from the visible and near UV emission of chromospheric origin as well as from H free-bound continuum. We did this exercise for all 5 sets of flares and the deduced parameters of the blackbody radiation are presented in table \ref{tab_nrj1}. The contribution of the WL continuum is similar in all cases, i.e. around 70\%. The estimated flaring areas are consistent with WLF observations \citep[e.g.][]{Hudson:2006aa} and increase with the flare magnitude. Finally let's also note that other emission mechanisms with similar spectral distribution should lead to similar estimates of the flare energy contained in the white-light and near UV continuum.\\ 
%
%
\subsection{The case of the X17 flare on 28 October 2003\label{sec_28oct}}
\begin{figure}[!t] 
    \begin{center}
  \includegraphics[width=0.5\textwidth]{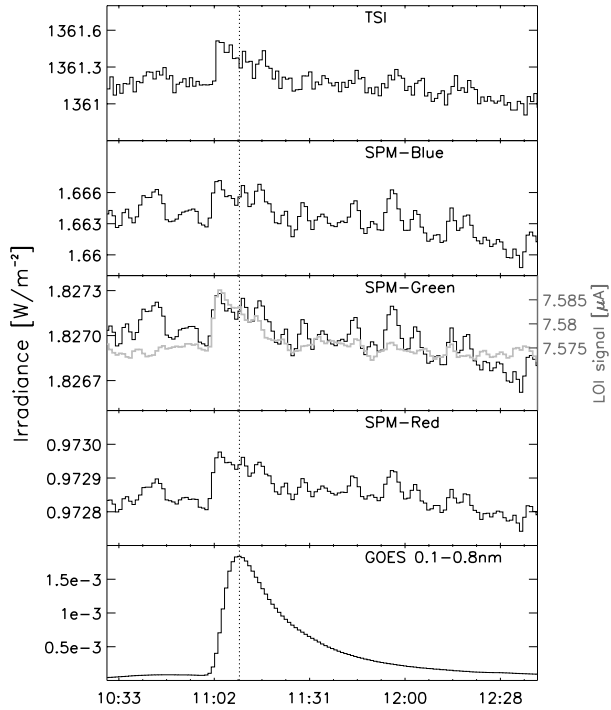}
    \end{center}
    \caption{{\bf Flare light curves for the October 28, 2003 flare} Single flare light curves for the X17 flare of October 28, 2003. The grey curve in the 3$^{rd}$ panel is the flux coming from a region of $\sim$900$arcsec^{2}$ that includes the flare site and observed by the VIRGO/LOI instrument \citep{1997SoPh..170...27A} in the same passband than SPM/Green. SEM and GOES 0.01-0.5nm are not shown as they are polluted by saturation and particle effects. \label{fig_28oct}}
\end{figure}
 The X17 flare thAT occurred on October 28, 2003 is a very large flare and the only one that has been unambiguously detected in TSI \citep{2004GeoRL..3110802W}. Fig.\ref{fig_28oct} shows irradiance light curves for this flare and confirms this finding by showing the flare signature in the SOHO/VIRGO instrument, while Woods et al. used the TIM instrument onboard SORCE. The increase in VIRGO is 264ppm while it is of 268ppm in the TIM measurements, i.e. in very good agreement. It also shows for the first time A clear white-light signature in Sun-as-a-star observations of a single flare using the VIRGO/SPM channels (relative increase of 267ppm, 191ppm, and 176ppm respectively for the blue, green, and red channel). Note that the visible light and TSI peak about 5 minutes before SXR, confirming the importance of the impulsive phase.\\
Following the analysis of the previous section, we find that the ratio $I_{blue}/I_{green}$ is 1.28 which corresponds to a blackbody temperature of 8545K, relatively lower than in the average cases of the previous sections, although still consistent. The signal-to-noise ratio is lower in this single event than in the average cases but this somehow gives an idea of the uncertainty on the determination of the blackbody temperature and we should speak of temperature {\it roughly around} 9000K, that does not change very much with the amplitude of the flare. Matching the observed blue signal gives a flaring area of $\sim$130 $arcsec^{2}$; this is much larger than areas shown in tab.\ref{tab_nrj1} but it should not be so surprising for such a flare; it also agrees with white-light observations from the Global Oscillation Network Group (GONG) -see fig.4 and fig.7 of \cite{2009SoPh..258...31M}). Integrating then a blackbody spectrum at 8545K over this flaring area, we find that the blackbody spectrum accounts for 64\% of the total energy, in good agreement with what we found for the average cases in table \ref{tab_nrj1}.

%
\section{Conclusions\label{sec_conclu}}
 In this study, we identify and analyze for the first time the visible light produced by flares in Sun-as-a-star observations. We use a superposed epoch analysis to show that visible emission is present on average for all flares from X-ray class X to C, that it occurs mainly during the impulsive phase and must be considered as continuum emission, i.e. white-light flares. Analyzing the intensity ratio of the SPM blue and green channels, we found this emission to be consistent with a blackbody spectrum near 9000K. We next match the increase observed in the blue channel and deduce flaring areas that are consistent with previous observations. Using these results, we compute the total energy contained in the continuum and find it to represent about 70\% of the total energy radiated by flares. This study shows that that the white light continuum is ubiquitous in flares and that it represents about 2 thirds of the energy radiated by the flares. Furthermore, we reveal the existence of white light signature in solar flux for the single X17 flare that occurred on October 28, 2003 and find similar energy distribution.  \\
These results show the very large predominance of the lower atmosphere with respect to the corona in freeing the flare energy that is initially stored in the solar magnetic field, and put constraints on models. Additionally, since each flare releases most of its energy in white light, and since it exists a continuum of flares, we can wonder if the visible emission released by flare can contribute to the variations of the TSI. We plan to address this issue in a forthcoming study.

%
\section{acknowledgements}
This work has received funding from the European Community's Seventh Framework Programme (FP7/2007-2013) under the grant agreement n° 218816 (SOTERIA project, www.soteria-space.eu). The author thanks T. Appourchaux for the VIRGO/LOI data, C. Wehrli for providing the SPM response functions, T. Dudok de Wit for useful discussions, and one anonymous referee for useful comments and suggestions.
%
%
\bibliographystyle{plainnat} 
\bibliography{Kretzschmar_SunasAStar_Flare_Preprint} 

\end{document}